\documentclass[pre,twocolumn,showpacs,superscriptaddress] {revtex4}

\usepackage{amsmath,amssymb}
\usepackage{graphics,graphicx}

\newcommand{\cu}
{\affiliation{Department of Physics, University of Calcutta, 92 APC Road, Kolkata 700009, India }}

\begin{document}

\title
{Susceptible-Infected-Recovered model on Euclidean network}

\author{Abdul Khaleque}%
\email[Email: ]{7187ak@gmail.com}
\cu
\author{Parongama Sen}%
\email[Email: ]{psphy@caluniv.ac.in}
\cu

\begin{abstract}


We consider the Susceptible-Infected-Recovered (SIR) epidemic model on a 
Euclidean network in one dimension in which nodes at a distance $l$ are
connected with probability $P(l) \propto l^{-\delta }$ in addition
to nearest neighbors. The topology of the network changes as $\delta$ is varied and its effect on the SIR model is studied.
$R(t)$, the recovered  fraction of population up to time $t$,  and $\tau$, the total duration of the epidemic are calculated for 
different values of   the infection probability $q$  and $\delta$.
A threshold behavior is observed for all $\delta$ up to $\delta \approx 2.0$; 
above the threshold value $q = q_c$,  the saturation value $R_{sat}$ attains a  finite value.  Both  $R_{sat}$ and $\tau $  show scaling behavior
in a finite system of size $N$;   
$R_{sat} \sim N^{-\beta/{\tilde{\nu}}} g_1[(q-q_c)N^{1/{\tilde {\nu}}}]$ and 
$\tau \sim N^{\mu/{\tilde{\nu}}} g_2[(q-q_c)N^{1/\tilde{\nu}}]$. 
 $q_c$ is constant for $0 \leq \delta < 1$ 
 and increases with $\delta$ for $1<\delta\lesssim 2$.
Mean field behavior is seen up to $\delta \approx 1.3$; weak dependence on $\delta$ is observed beyond this value of
$\delta$. 
The distribution of the outbreak sizes is  also estimated and found  to be unimodal for $q < q_c$ and bimodal for $q > q_c$.
 The results are  compared to static percolation phenomena
and also to mean field results for finite systems.
Discussions on the properties of the Euclidean network are made in the light of the present  results.


\end{abstract}
\maketitle


\section{Introduction}

Studies of  familiar static and dynamic phenomena  on complex networks have led to 
some surprises 
\cite{newman1,doro_rmp,barrat} in recent times.
For example, the Ising model which in one dimension does
not have a phase transition, showed not only the existence of a 
phase transition but also  that it  occurs with mean field criticality \cite{doro_rmp,ising_WS,gitterman,herrero,choi}
on the Watts-Strogatz (WS) \cite{WS} type   network. On the scale-free network also, it showed 
a behavior not encountered usually on regular lattices;
the transition temperature showed a logarithmic increase with the system size \cite{herrero2}.
Among the well studied dynamical   phenomena on networks  \cite{barrat} 
are 
opinion dynamics models \cite{castellano},   disease and damage propagation \cite{vesp1,vesp2,newman,moreno}, 
synchronization of coupled oscillations \cite{gade,doro_rmp},   
zero temperature quench of Ising and other spin models \cite{svenson,hagg,boyer,das,soham_sen},
etc. Just like
the static results on networks, there have been special features of the
dynamical phenomena as well. On complex networks, zero temperature quenching of Ising model 
shows freezing even in a one dimensional network with additional  
random  links. Another example is the voter model: its dynamics  can be conceived in different
manners on networks leading to the conservation 
of different physical quantities. However,  on regular lattices these 
different dynamical rules are equivalent \cite{castellano}.
%

A standard model for epidemic spreading is the  Susceptible-Infected-Recovered (SIR) model
where individuals can be in three possible states; susceptible: who are liable to be infected,
recovered: those who contracted the disease but are now recovered and immunized, 
infected: people who are suffering from the disease and can infect others.
A very well known result for the  SIR model is that it
 is like 
 a dynamical percolation problem and its critical behavior
coincides with that of static isotropic percolation on regular lattices \cite{grass}. On the other hand, another model 
for disease spreading, namely the Susceptible-Infected-Susceptible
(SIS) is identical to a  directed percolation problem. 
Percolation 
phenomena has    also been studied extensively on  complex networks 
(earliest works appear in  \cite{new-watts,cohen2000,moore} and comprehensive reviews 
are available in  \cite{doro_rmp,barrat})
and it is possible to compare the two phenomena of epidemic spreading and percolation on complex networks as well.  

Although extensive research work on epidemic spreading on scale free and WS networks have been made,
it may be noted  that social connections are neither scale free nor 
like those considered in the WS network. In fact, many social
networks have a spatial dependence in the connection probability of the nodes
\cite{barthelemy}. We thus consider SIR on a spatial
model  
in which 
 random long range 
links between nodes at a Euclidean distance $l$ are 
added with probability $P(l) \propto l^{-\delta}$.
Nearest neighbor links are always present in this model. 
In \cite{warren1,warren2},  a similar model in two dimensions was considered,
but the probability $P(l)$  was essentially dictated 
by the heterogeneity of the degree distribution,  and only short range 
links followed the power law distribution while  long range  
links were added randomly.

Static properties of the  network considered 
in this paper  are quite well studied \cite{sen-review,goswami}. The network behaves as a  small world network for 
$\delta < 1$ and as a regular one dimensional lattice for $\delta > 2$. Some
ambiguities remain regarding its behavior when  $1 < \delta <2$.  While some results 
suggest the network is still a small world here, other works claim that it has a finite lattice-like behavior. 
 Dynamic phenomena  like zero temperature 
coarsening of Ising model and searching  
have also been studied  on this network \cite{zhu,soham_sen}.


The aim is to locate  the infection threshold values as a function of $\delta$ 
and find out the critical behavior for SIR on the Euclidean network. This may 
also help in understanding the nature of the network in the controversial region $1 < \delta < 2$ by  considering a  dynamical process. 
We have also studied the way the recovered  fraction grows in time as well as the total  
duration of the entire epidemic spreading process.  

%

In section II, we describe  the model and method briefly and follow it up with the results in section III. 
In section IV, we discuss a simple model (mean field type) to make some comparisons. 
In section V, the results are discussed and compared to earlier studies made on the same network.
In the last section concluding remarks are made.
%


\section{MODEL AND METHODS}
The network is generated in the following way: in  a system of $N$ nodes 
placed on the sites of  a one dimensional lattice, all nearest neighbors  are first  
connected.
Additional long range links (one per node on an average)  are  established  
at a distance $l> 1$. This is done in the following way: two nodes are selected randomly; if they are not connected already,  
a connection is  established 
with a probability $P(l) \propto l^{-\delta}$ ($l$ is the  distance  separating the
nodes). 
The process is stopped as soon as $N/2$ links have been formed this way. The average degree of the  
nodes is therefore three. 

In a homogeneous system, the SIR model can be described in terms of the densities
of susceptible, infected and recovered nodes, $S(t)$, $I(t)$ and $R(t)$, respectively, as function of time. These three densities are related through the normalization 
condition: 
\begin{equation}
S(t)+I(t)+R(t)=1,
\label{norm}
\end{equation}
and they obey the following system of differential equations \cite{marro-dick}:
\begin{eqnarray}
\label{diffeq}
\frac{dS}{dt} &=& -q(k-1)IS, \\
\frac{dI}{dt}& =&-{\mu}I +q(k-1)IS, \\
\frac{dR}{dt}& =&{\mu}I. 
\end{eqnarray}
These equations can be interpreted as follows: infected nodes become recovered at a rate $\mu$, while susceptible nodes become infected at a rate 
proportional to both the densities of infected and susceptible nodes. Here, $q$ is the infection rate and $k$ is the number of contacts
per unit time.

In the simulation, 
systems with size $N<2^{14}$ have been taken. 
  Time is discretized and only the infection rate $q$ is used as a parameter; 
infected people recover within one unit of time and can infect susceptible individuals with probability $q$ 
connected directly to them within the same time scale. 
Initially all nodes are susceptible, one arbitrary node is chosen and taken to be infected.
 For the same network, 600 such choices 
have been taken and quantities are averaged. A secondary averaging is made by considering 100 different 
network configurations. Dynamics takes place in parallel; all the  individuals revise
their epidemiological state in one time step.  Periodic boundary condition has been used in the simulation.


\section{Results}

\subsection{Static results}

\subsubsection{Fraction of recovered population}

We calculate the fraction $R(t)$ of recovered  nodes as a function of time. This means the total fraction of nodes
who are recovered till the time of measurement.  It is also identical to the fraction of the population who have been infected at some point of time in the past. 
$R(t)$  reaches a saturation value $R_{sat} \leq 1$
which depends on both $q$ and $\delta$.  We use synchronous dynamics to update the state of the nodes.
The total duration $\tau$ of the epidemic is also estimated. 


\begin{figure}
\includegraphics[width=2.9cm,angle=270]{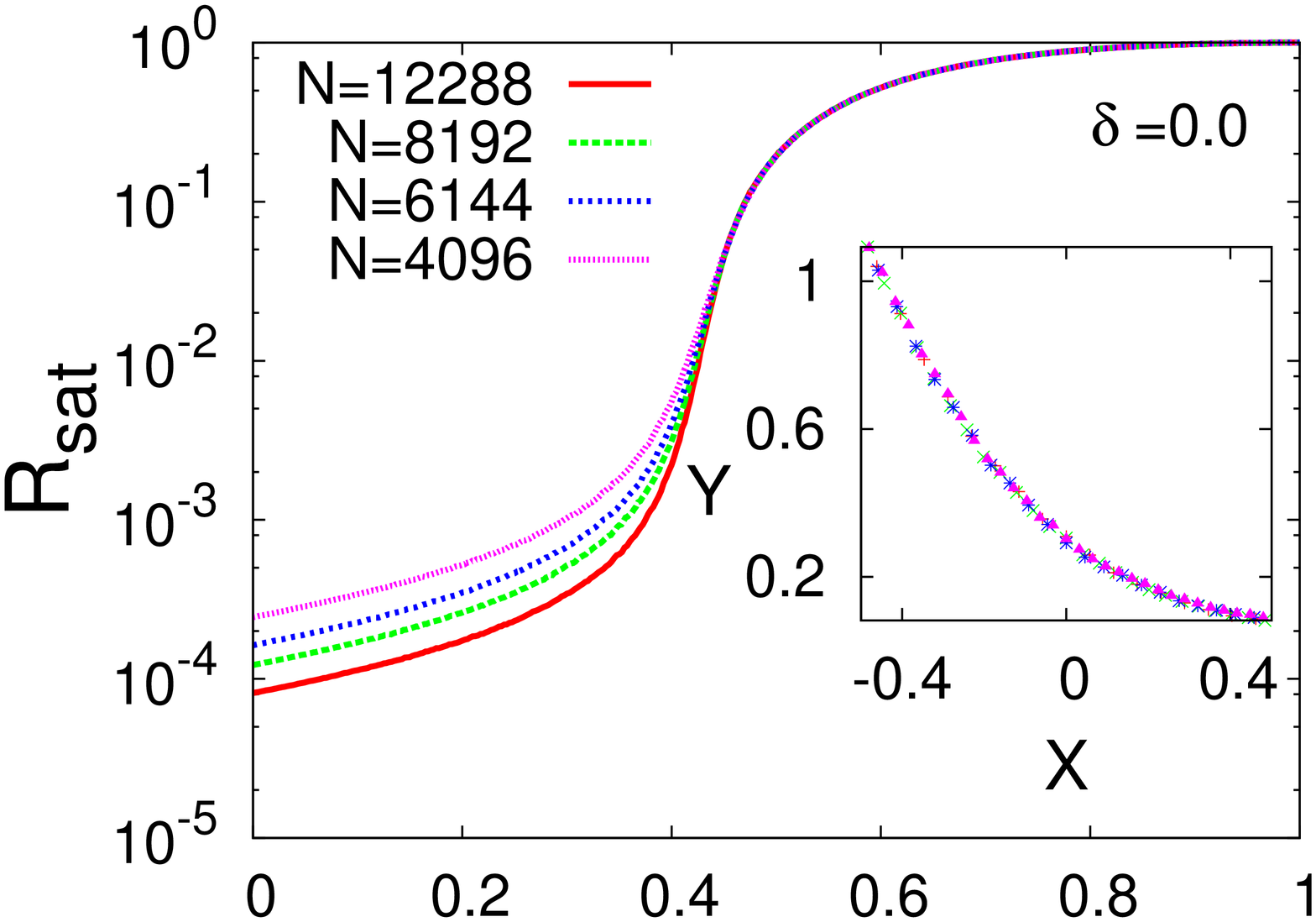}
\includegraphics[width=2.9cm,angle=270]{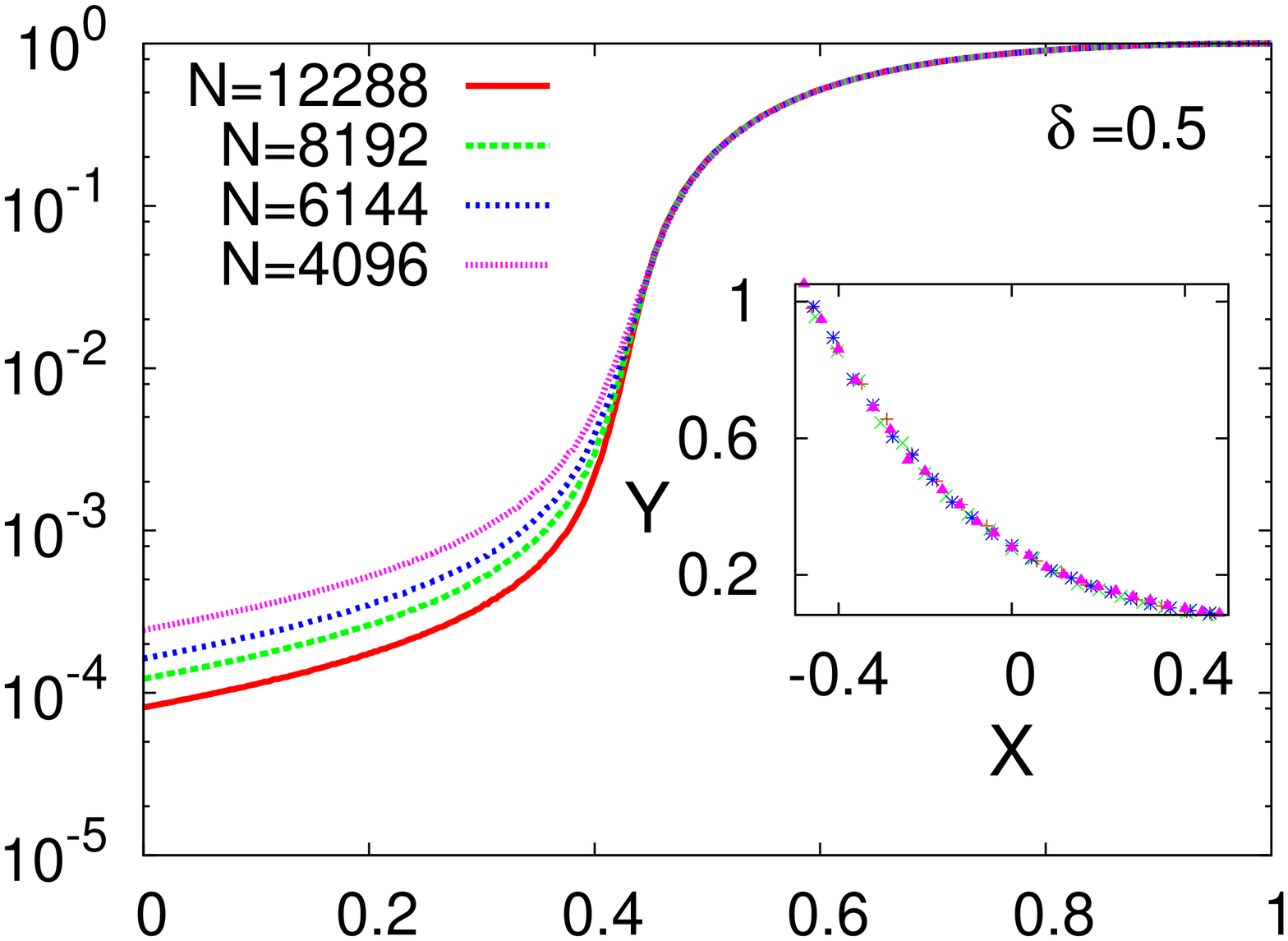}

\includegraphics[width=2.9cm,angle=270]{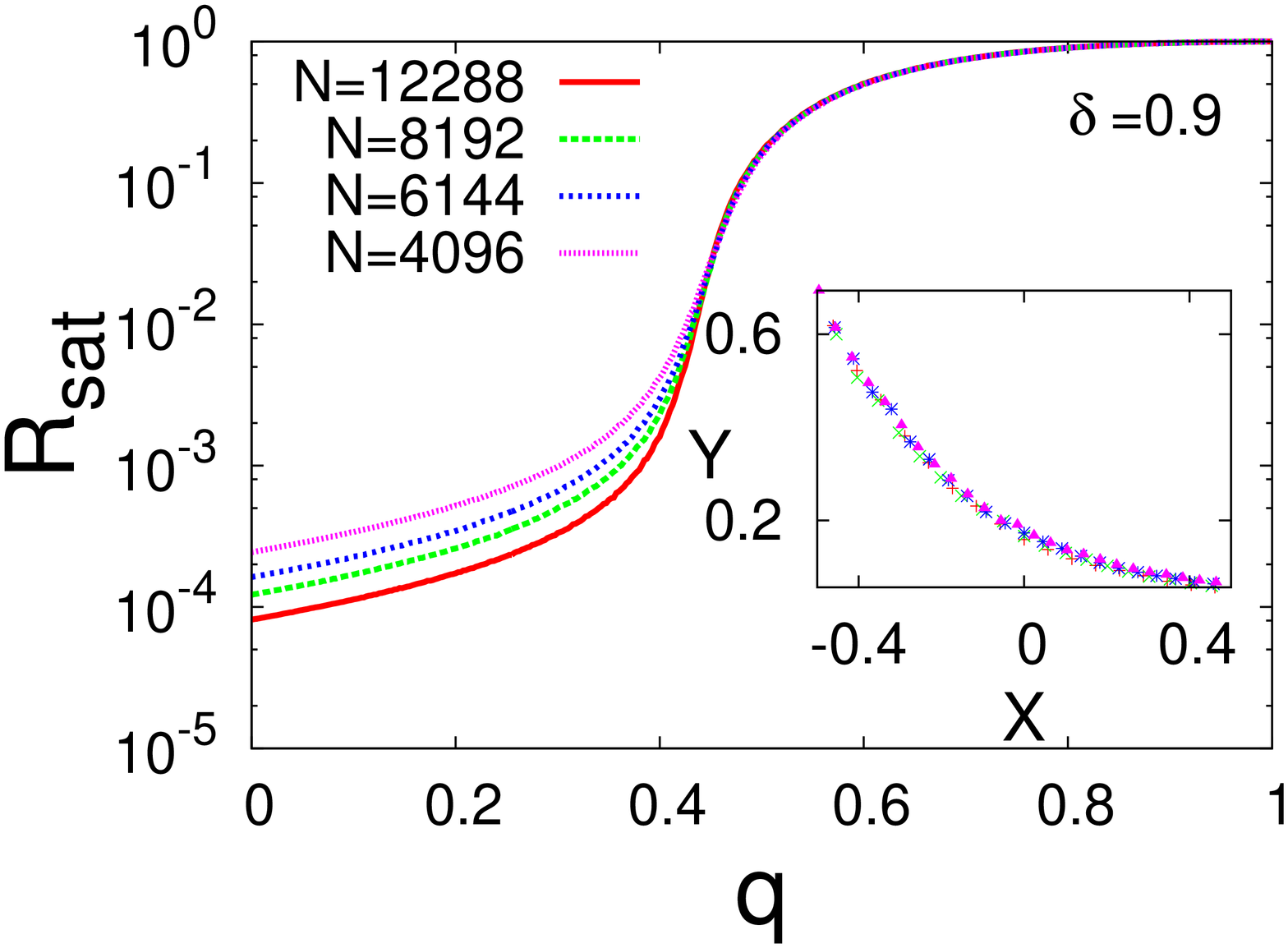}
\includegraphics[width=2.9cm,angle=270]{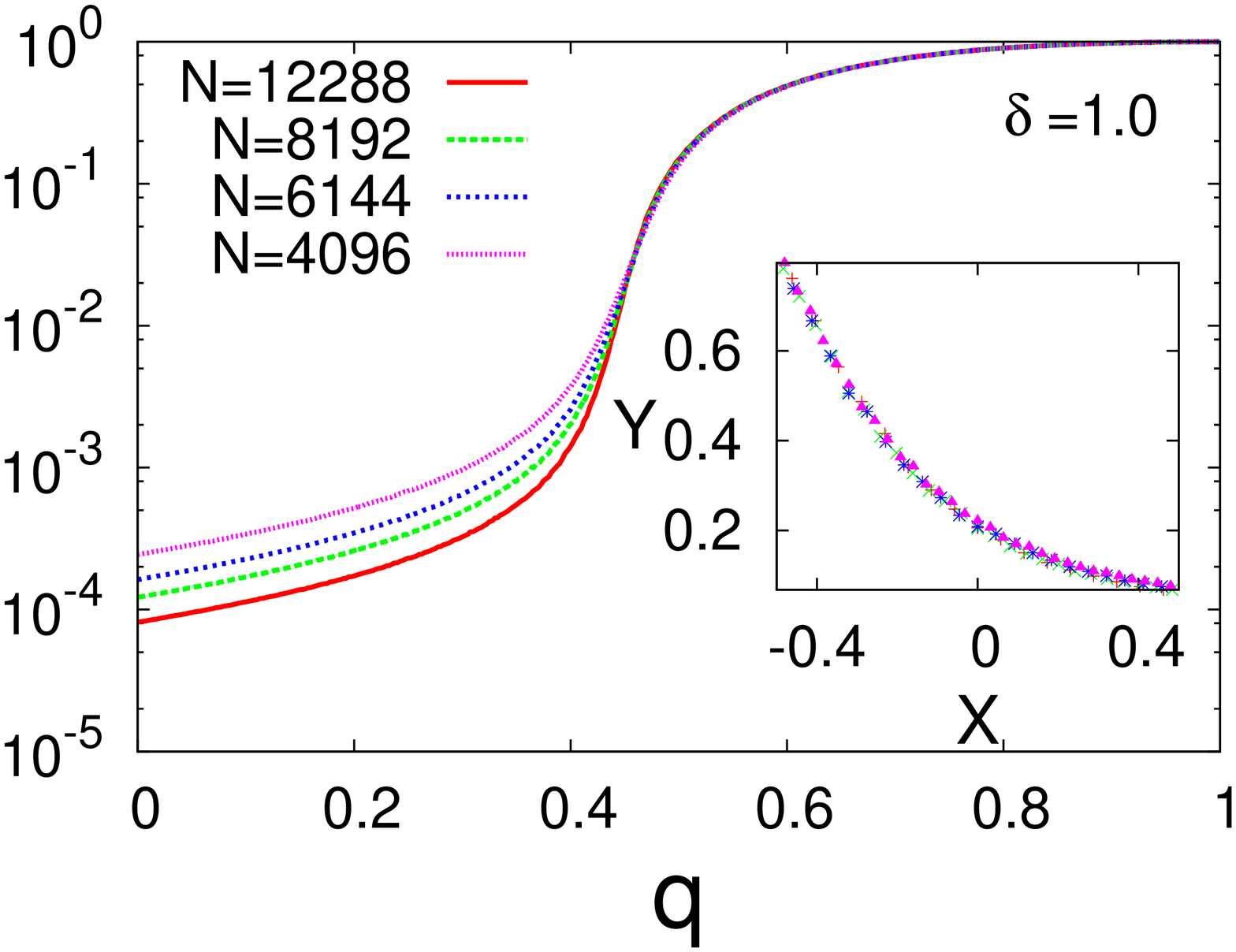}
\caption{(Color online) Variation of $R_{sat}$ with $q$  for different values of $\delta=0.0, 0.5, 0.9$ and $1.0$.
Insets show the data collapse where $Y=R_{sat}N^{\beta/\tilde{\nu}}$ has been plotted against $X=(q-q_c)N^{1/\tilde{\nu}}.$}

\label{fsat}
\end{figure}

\begin{figure}
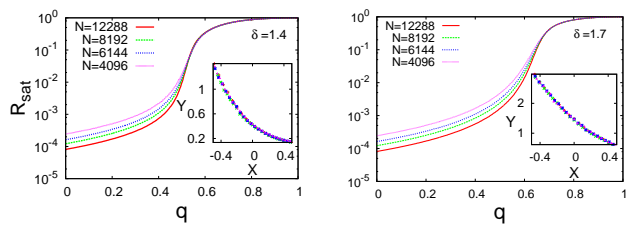

\includegraphics[width=2.9cm,angle=270]{data_collapse1.4.eps}
\includegraphics[width=2.9cm,angle=270]{data_collapse1.7.eps}

\caption{(Color online) Variation of $R_{sat}$ with $q$  for different values of $\delta=1.4$ and $1.7$.
Insets show the data collapse where $Y=R_{sat}N^{\beta/\tilde{\nu}}$ has been plotted against $X=(q-q_c)N^{1/\tilde{\nu}}.$}

\label{fsat1}
\end{figure}

As a function of the infection probability $q$, $R_{sat}$ is plotted for different values of $\delta$ (Fig. \ref{fsat}). 
From the knowledge that SIR shows the same criticality as the isotropic percolation problem,
it can be expected that a continuous phase transition takes place here. 
Since it is a finite size system, one can use the conventional finite size scaling method to 
estimate the  critical (threshold) value  $q_c$ and the 
associated critical exponents. $R_{sat}$ vanishes below $q_c$ in the thermodynamic limit and should have a scaling
form similar to magnetization. Therefore, the  following finite size scaling form for  $R_{sat}$ is used: 
\begin{equation}
R_{sat} \propto N^{-\beta/\tilde{\nu}} g_1((q-q_c)N^{1/\tilde{\nu}}),
\end{equation}  
where $g_1$ is a scaling function. 
We obtain data collapse with appropriate values of $\beta$, $q_c$ and $\tilde{\nu}$ (Fig. \ref{fsat} insets).
Let us first discuss the region $0 \leq \delta < 1$ where $q_c \simeq 0.430$ does not show  any appreciable dependence on $\delta$.
For $\delta = 0$, the results should be comparable to a (addition type) Watts-Strogatz type network.
The percolation threshold is known to be 0.401 there  \cite{moore}. Our result for $q_c$ is slightly higher
which could be   because  the 
average degree is different and also the fact that  nearest neighbor links are always present here such that
in comparison, long range neighbors are smaller in number.

Results for $0 \leq \delta<1$ show that the value of $\beta\simeq1$  and $\tilde \nu \simeq 3$ are also  unchanged
in this region. This is not surprising, this region is known to have  a small world behavior. Hence, mean field behavior is expected to be valid here and 
we find the value exponent $\beta =1$ indeed matches with the mean field value
for percolation. 
Assuming   
$\tilde {\nu} = \nu d$, where $d=6$ is the upper critical dimension \cite{herrero,arnab_ps,bcs,doro_rmp},
 one gets $\nu =1/2$ from the fact that $\tilde {\nu} \simeq  3$,  which also coincides
with the mean field value for percolation phenomenon. 

For $\delta > 1$, $q_c$ increases and approaches 1 at a value of $\delta $ close to 2. This is expected, as for $\delta >2.0$, the network behaves as a regular
network in one dimension where $q_c=1$.
The value of the exponent $\tilde{\nu}$ however, appears to be unchanged while $\beta$ shows 
a variation with $\delta$, although not very strong. 
$\beta$ starts deviating from the mean field value at around $\delta = 1.4$. 

\begin{figure}
\includegraphics[width=2.9cm,angle=270]{time_collapse0.0.eps}
\includegraphics[width=2.9cm,angle=270]{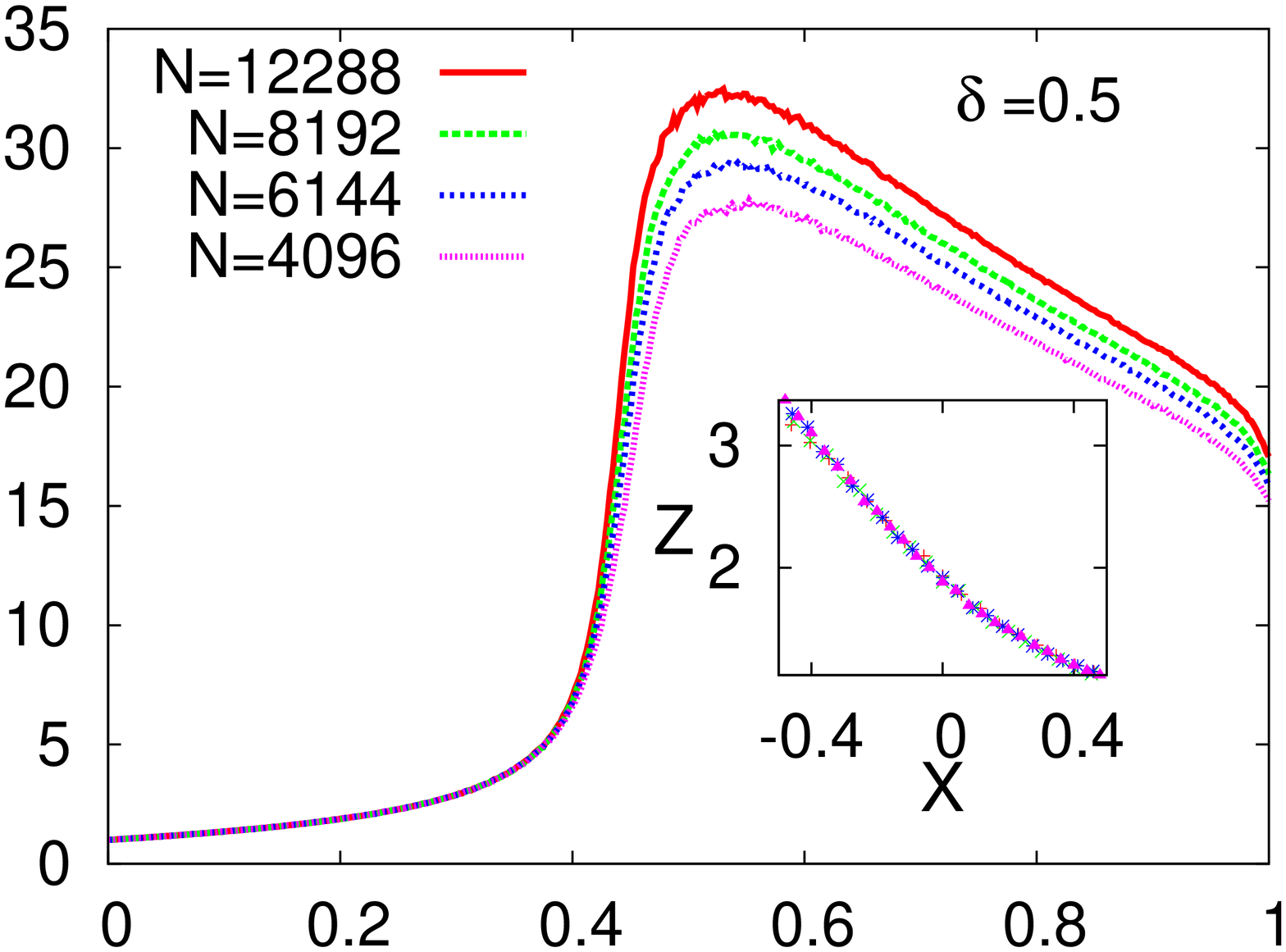}

\includegraphics[width=2.9cm,angle=270]{time_collapse0.9.eps}
\includegraphics[width=2.9cm,angle=270]{time_collapse1.0.eps}
\caption{(Color online) Variation of $\tau$ with $q$  for different values of $\delta=0.0, 0.5, 0.9$ and $1.0$.
Insets show the data collapse where $Z={\tau}N^{-\mu/\tilde{\nu}}$ has been plotted against $X=(q-q_c)N^{1/\tilde{\nu}}$.}
\label{times}
\end{figure}

The total duration $\tau$ also shows a dependence on $q$; its peak value increases with the system size (Fig. \ref{times}). This data 
 is also analyzed by finite size scaling 
assuming the scaling form
\begin{equation}
\tau \propto N^{\mu/\tilde {\nu}} g_2((q-q_c)N^{1/\tilde {\nu}}),
\end{equation} 
where $g_2$ is another scaling function. 
$\tilde \nu$ turns out to be very  close to 3.0 from the above analysis as well for all $\delta$ values, 
while $\mu$ shows a  dependence on $\delta$
for  $\delta  \geq  1.3$.

\begin{figure}
\includegraphics[width=2.9cm,angle=270]{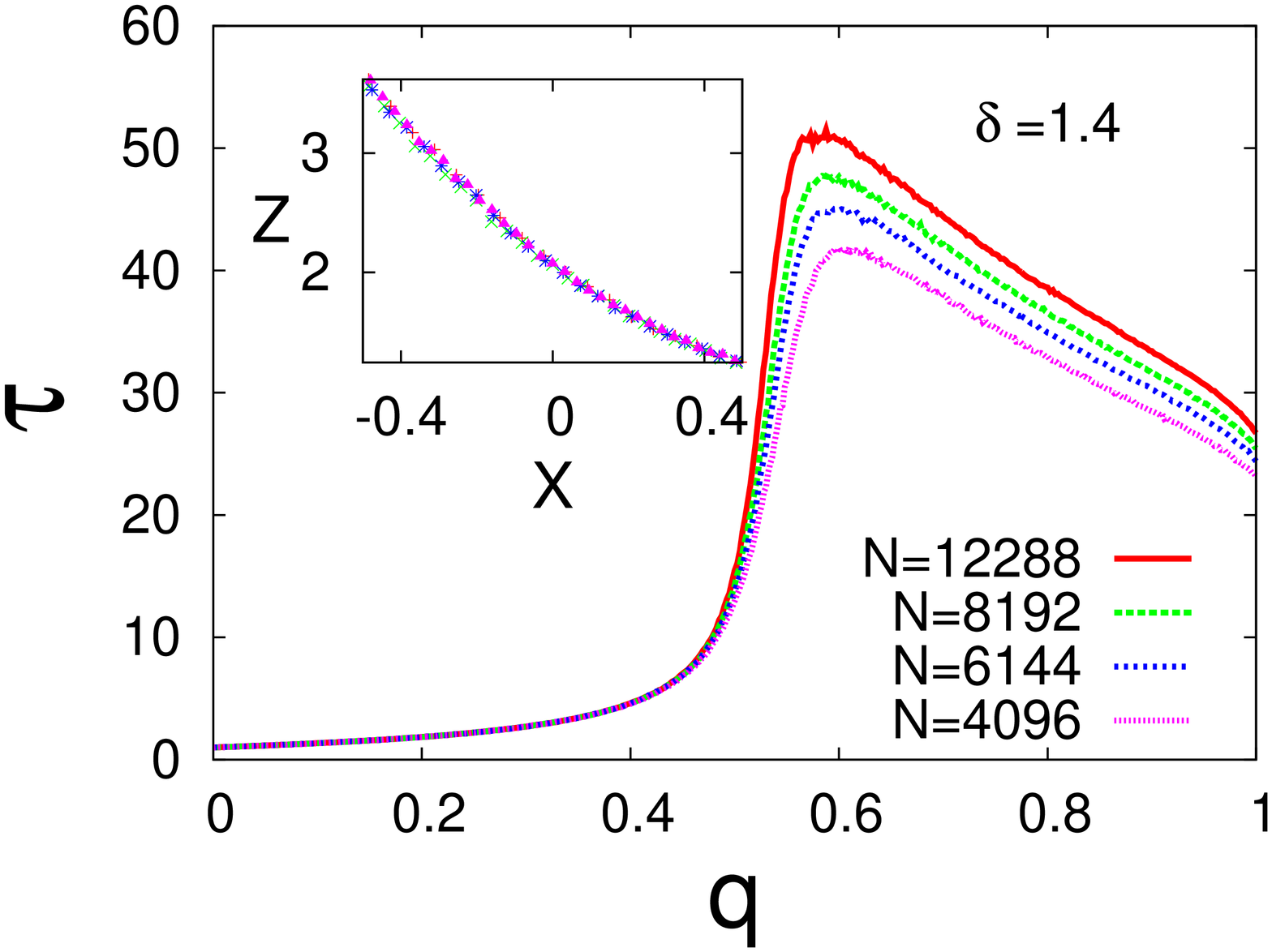}
\includegraphics[width=2.9cm,angle=270]{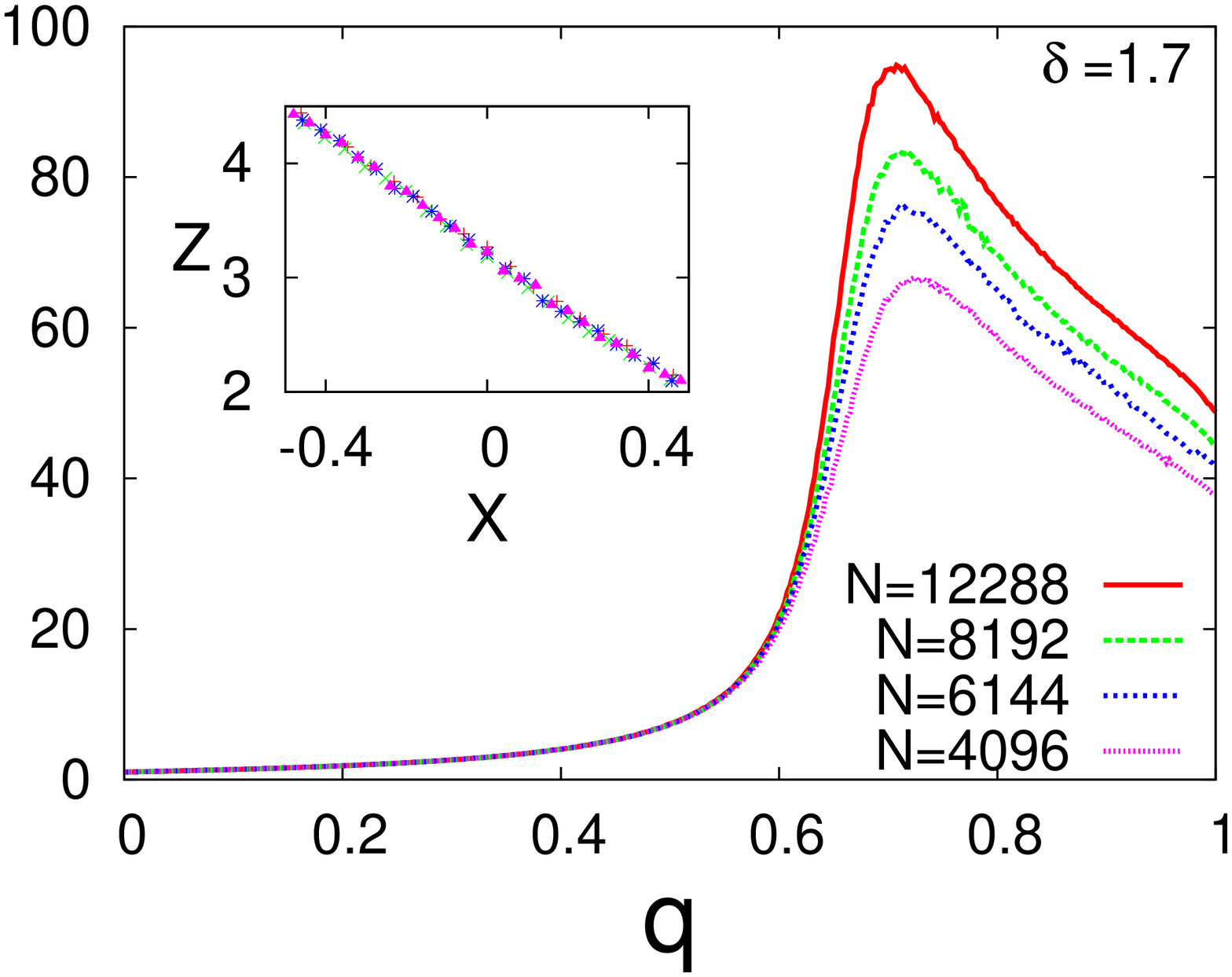}
\caption{(Color online) Variation of $\tau$ with $q$  for different values of $\delta=1.4$ and $1.7$.
Insets show the data collapse where $Z={\tau}N^{-\mu/\tilde{\nu}}$ has been plotted against $X=(q-q_c)N^{1/\tilde{\nu}}$.}
\label{times1}
\end{figure}

\begin{figure}
\includegraphics[width=6.5cm]{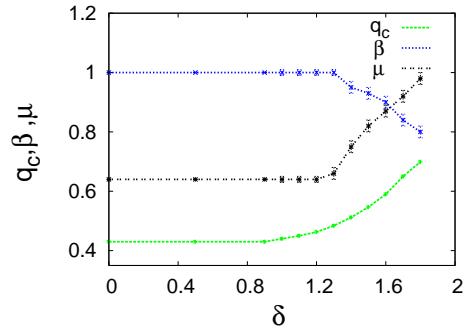}
\caption{(Color online) The exponents and the transition point shown against $\delta$.}
\label{expo}
\end{figure}

We show the values of the threshold values $q_c$, and the exponents $\mu$ and $\beta$ as  functions of $\delta$ in 
Fig. \ref{expo}. Here $q_c$ shows variation with $\delta$  for $\delta > 1.0$, the values of the exponents start differing from the mean 
field value  at $\delta >  1.3 $. $q_c$ appears to reach 1 slightly above $\delta=2.0$,
we have checked that for $\delta=2.2$, $q_c$ is very close to 1.

If we put $q=q_c$, the variation of $\tau$ with $N$ 
shows that   
it diverges algebraically  
with $N$ sublinearly in the entire region $0 \leq \delta \leq 2$. 
         This is interesting: $\tau$ 
         may be regarded as the  minimum  number of steps  connecting two individuals for exactly $q = 1$ 
           in the network, i.e., it is comparable to the 
 diameter of the network. However,
one might expect that for $q \geq q_c$, when spanning occurs, a finite number  of the nodes can be reached by
the infection procedure and hence $\tau$ gives an estimate of the  average number of steps connecting two individuals for the entire network 
(which may not necessarily be the minimum path). 
 We know that in the small world region,
        the diameter scales as $\log(N)$ but here we get a different scaling for $\tau$.  
We will discuss this point in section IV and V. 

\subsubsection{Distribution of the outbreak size}

\begin{figure}
\includegraphics[width=6.5cm,angle=0]{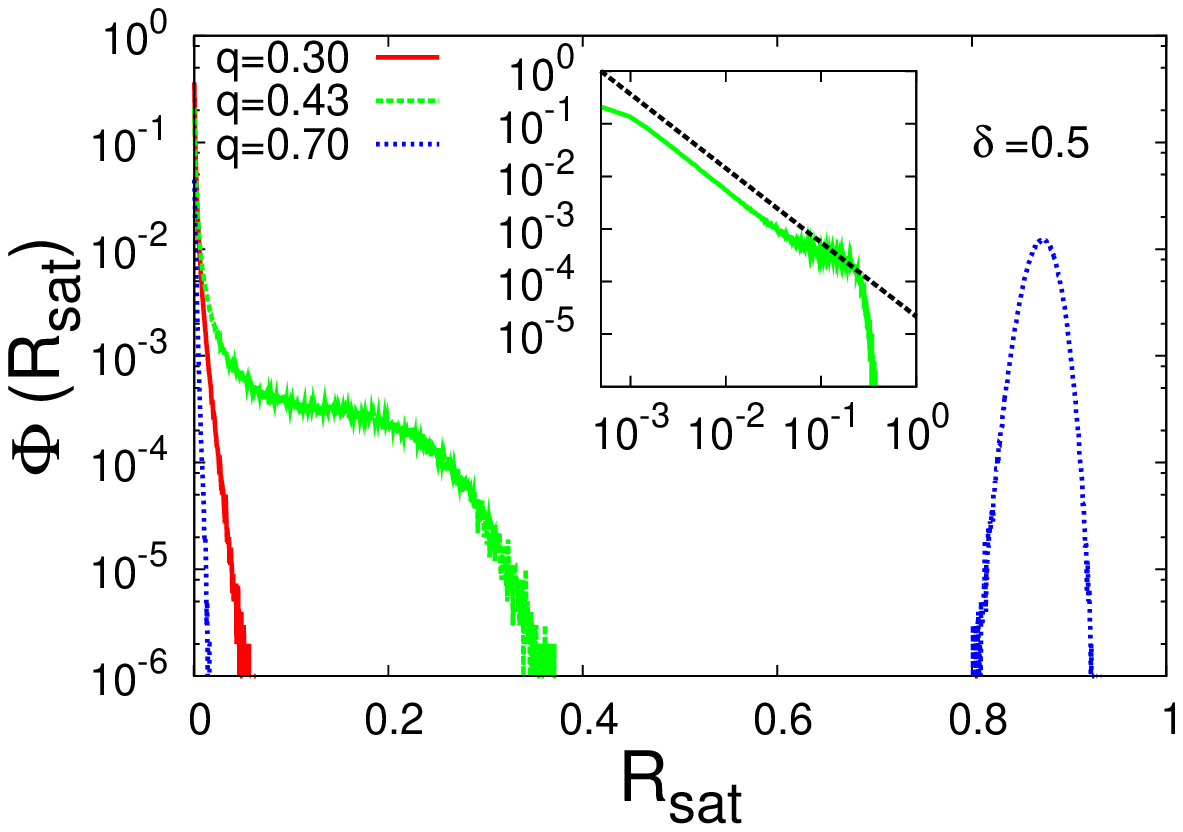}
\includegraphics[width=6.5cm,angle=0]{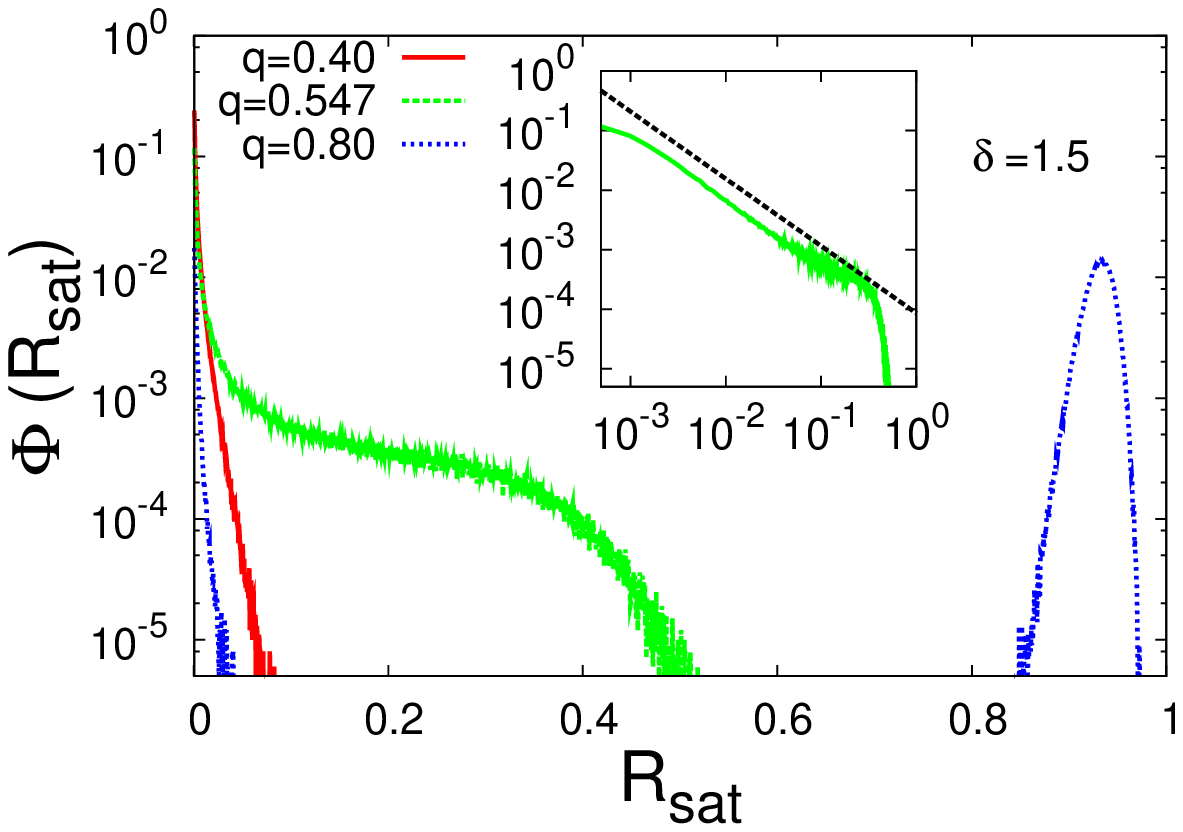}
\caption{(Color online) Frequency distributions for outbreak size $R_{sat}$  are shown for $N=2048$ for two values of $\delta$.
Data for $q < q_c, q > q_c$ and $q = q_c$ are shown. For $q< q_c$ the distribution is unimodal while for $q > q_c$, 
it is bimodal with two peaks occurring at $R_{sat} \sim 1/N$ and  $R_{sat} \sim O(1)$ (see text). Insets show that the data at $q = q_c$ has a power law variation. 
}
\label{dist}
\end{figure}

In the discussion in the preceding subsection, results for the average value of the outbreak size $R_{sat}$ has been reported.
We have also studied the distribution $\Phi(R_{sat})$  of  $R_{sat}$  which shows that it is bimodal in nature for $q>q_c$ and unimodal for $q < q_c$  as 
found in \cite{zanette,gallos}. 
$\Phi(R_{sat})$ has a peak near $\frac {1}{N}$ for all values of $q$ which appears  for  the cases when the initially infected single node cannot transmit the disease to anyone else. 
The peak value at $1/N$ decreases with $q$ as expected. For $q>q_c$ the $\Phi(R_{sat})$ has a secondary peak at a larger value of $R_{sat}\sim O(1)$ (Fig. \ref{dist}). It may be noted that 
the peak values at $R_{sat}\sim\frac{1}{N}$ and $R_{sat}\sim O(1)$ are comparable for $q > q_c$ which makes the average value of $R_{sat}$ a meaningful quantity. 
At $q_c$, $\Phi(R_{sat})$ varies continuously and the tail of $\Phi(R_{sat})$ has power law decay as shown in the inset of Fig \ref{dist}. The associated exponent varies with $\delta$, for example, 
it is  $\sim  1.41$ for 
$\delta = 0.5$ and $\sim 1.13$ for $\delta = 1.5$.  

Such a bimodal behavior of distribution of outbreak sizes, which occurs as a single agent is assumed to be infected initially, is consistent with 
empirical data of disease spreading analyzed by Watts et al \cite{watts}.

\subsection{Dynamical results}

\begin{figure}
\includegraphics[width=4.3cm,angle=270]{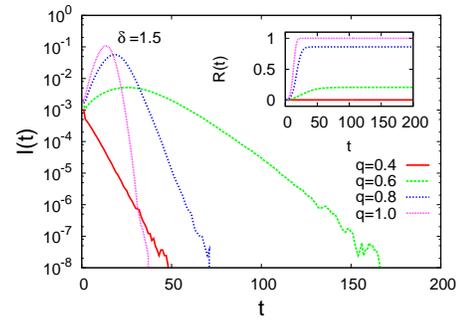}
\caption{(Color online) The  infected fraction of nodes shown against time for $\delta = 1.5$ for a system of size $N = 2048$.
There is a peak at $t_p$ for $q > q_c$. Inset shows the variation of total recovered fraction of nodes with time for $\delta = 1.5$.}
\label{new-inf}
\end{figure}

The growing density of 
      recovered    population $R(t)$ has been plotted in the inset of Fig. \ref{new-inf}.  The data shows the expected saturation and a very fast growth at initial times suggesting an exponential behavior at early times \cite{heth}. 
 It is found that the  numerical data for $R(t)$ can be fitted to the form:
\begin{equation}
R(t) = \frac{a\exp(t/T)}{1+c\exp(t/T)}-\frac{a}{1+c},
\label{fit}
\end{equation}
where $a$, $c$ and $T$  depends on $q$ and  $\delta$. The boundary condition 
assumed in the fitting
 is $R(0)=0$.

The fraction of   nodes  infected at time $t$,  
$I(t) = R(t+1) -R(t)$  is    
plotted in Fig. \ref {new-inf}.  
An initial growth and  a  peak value occurs  at time $t = t_p$ only for  $q > q_c$.  For $q < q_c$, 
one gets a decaying behaviour right from $t=0$. Such a decaying   behaviour
can occur  if secondary infections are less  than primary infections.
Hence approximately,  
\begin{equation}
(k-1)^2q^2 <  (k-1)q
\end{equation}
which gives
\begin{equation}
q <  1/(k-1)=q_c.
\end{equation}
Hence this argument can explain the absence of the peak for $ q < q_c$. The fact that the recovered population is no longer susceptible
 has been ignored in this argument, 
but for  initial times, this will not matter when the recovered population is very small.
This is supported by  the data presented  in Fig \ref{dist} which show that for   $q<q_c$, $R_{sat}$  has very small values only.

\begin{figure}
\includegraphics[width=6.7cm,angle=0]{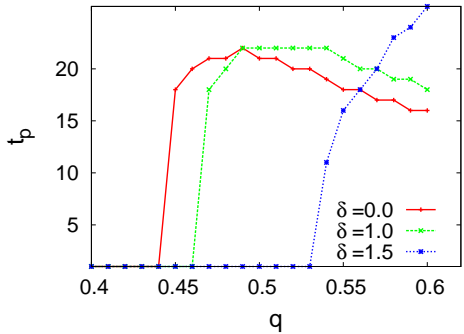}
\caption{(Color online) The time $t_p$ at which the peak value occurs against $q$ shows a sharp rise near $q_c$ (Data shown for $N=2048$).}
\label{peaks1}
\end{figure}

\begin{figure}
\includegraphics[width=6.8cm]{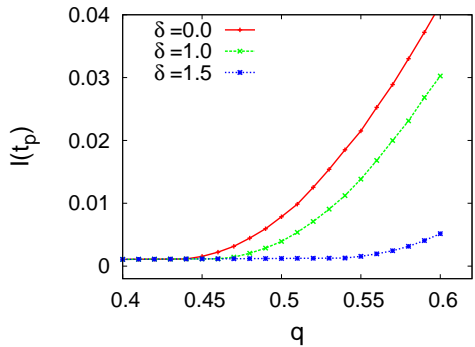}
\caption{(Color online) 
 The peak value of $I$  against $q$ shows a monotonic increase (Data shown for $N=2048$).} 
\label{peaks2}
\end{figure}

The variation of $t_p$ against $q$ shows  $t_p$  increases sharply 
 from zero close to the  
transition point $q_c$ (Fig. \ref{peaks1}). Thus one can get an independent estimate of $q_c$ 
from this study. One may also plot the peak value
 $I(t_p)$ as a function of $q$ which again shows an increase from zero  close to the transition point  (Fig. \ref{peaks2}). 

One can check how good is the fitting  (eq. \ref{fit}) 
by the following approach.
Assuming 
continuous time, $I(t)$  can be expressed as $\frac{dR}{dt}$. 
Let $F(t)=\frac{dR}{dt}$, then from eq. (\ref{fit}), 
\begin{equation}
F(t) = \frac{\frac{a}{T}\exp(t/T)}{(1+c\exp(t/T))^2}.
\end{equation}
 At large times, $F(t)$ should fall exponentially with $t$, $I(t)$ indeed shows such a behavior.    
 The value of $t=t_p$ where  $F(t)$  is maximum is found by solving the 
equation     
\begin{equation}
\frac{dF}{dt} = \frac{\frac{a}{T^2}\exp(t/\tau)-\frac{ac}{T^2}(\exp(t/\tau))^2}{(1+c\exp(t/T))^3}=0,
\end{equation}
which gives 
\begin{equation}
t_p = T\ln\left( \frac{1}{c} \right)\\.
\label{t_p}
\end{equation}
At $t_p = T\ln(\frac{1}{c})$, the value of $F(t)$ is \\
\begin{equation}
F(t_p)=\bigg(\frac{dR}{dt}\bigg)_{t_p}=\frac{a}{4cT}. 
\label{F_p}
\end{equation}
Putting the values of $a$, $c$ and $T$ obtained from the fitting, 
$t_p$ and $F(t_p)$ can be computed from eqs (\ref{t_p}) and (\ref{F_p}) and compared to the actual data.
It shows very good agreement showing the quality of the fit. The comparison is shown in Table I for a particular value of $q$.

\begin{center}
\begin{table}
\caption{$a,c$ and $T$ for three values of  $\delta$  and comparison of 
$t_p$ and $F(t_p)$ obtained from fitting and data($q=0.58$)}
\begin{tabular}{|p{.5cm}|p{.98cm}|p{.98cm}|p{.9cm}|p{.94cm}|p{.94cm}|p{1.3cm}|p{1.2cm}|}

\hline

$   \delta $   &   $a$         &$  c$  & $  T  $   &   $ t_p  $ &   $ t_p  $& $F{\times}10^{-3}$    & $F{\times}10^{-3} $  \\
              &  ${\times}10^{-3} $    &${\times}10^{-3} $      &         & (fit)  &   (data) & (fit)&  (data) \\    \hline
 0.0     &       3.69 &  7.95 &   3.47  &  16.78 & 17 & 33.3 & 33.0 \\  \hline

 1.0     &  5.34  & 12.8 & 4.47  &       19.48   & 19& 23.3  &23.1\\ \hline

 1.5     & 19.7  & 133 &  11.87  &       23.95    & 23 & 3.12 &  3.15 \\  \hline

\hline
\end{tabular}
\end{table}
\end{center}

\section{MEAN FIELD RESULTS FOR FINITE SIZES}

One can formulate a mean field type recursion relation with discrete time steps
for the SIR  model in which the average degree is $\langle k \rangle$ 
from eq. (\ref{diffeq}), 
\begin{equation}
 I(t)=q(\langle k \rangle-1)I(t-1)S(t-1).
\end{equation}

As $I(t)=R(t+1)-R(t)$ and $S = 1-R-I$, 
this gives,
\begin{equation}
R(t+1)-R(t) = q(\langle k \rangle-1)(R(t) - R(t-1)) (1-R(t)).
\end{equation}
The initial 
conditions are 
$R(0) = 1/N$ and $R(1) =R(0)+q\langle k \rangle R(0)(1-R(0))$.
 Obviously,
here one assumes that the neighbors to which the infection spreads can be anywhere.
We note that the system size enters through the initial conditions 
only, 
and it has been assumed that only one individual is infected in the beginning.
%
%

\begin{figure}
\includegraphics[width=5.4cm,angle=270]{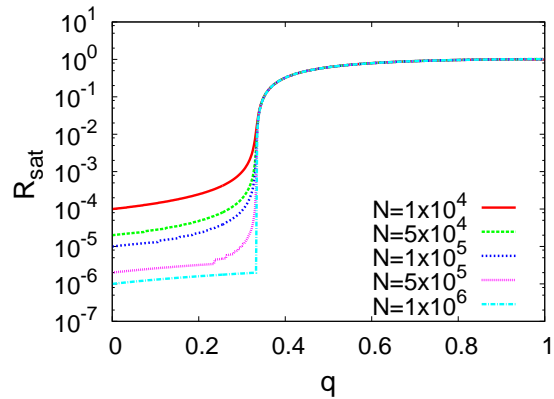}
\caption{(Color online) Variation of $R_{sat}$ with $q$ for mean field SIR model.}
\label{mf-fsat}
\end{figure}

\begin{figure}
\includegraphics[width=5.4cm,angle=270]{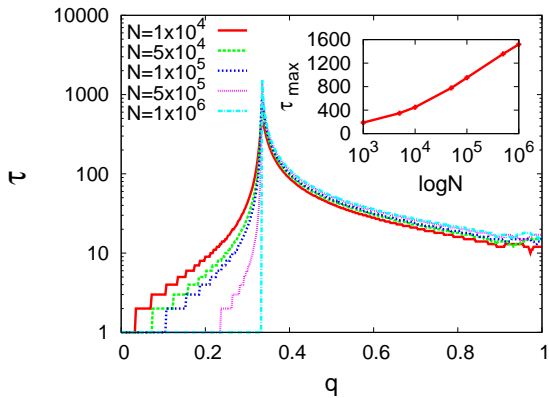}
\caption{(Color online) Variation of $\tau$ with $q$ for mean field SIR model. Inset shows the behavior of $\tau_{max}$ with system size for mean field SIR model.}
\label{mf-tau}
\end{figure}

$R(t)$ and $\tau$ are numerically estimated for different sizes using $\langle k \rangle=4$ and shown in Figs \ref{mf-fsat} and \ref{mf-tau}. The threshold value of $q$ is very close to the theoretical  
estimate $q_c = 1/(\langle k \rangle-1)=1/3$ \cite{moreno}.

These data however do not show any finite size scaling
behavior, i.e., a data collapse cannot be obtained. This is compatible with
the fact that for a mean field model, finite size scaling does not work \cite{brezin}. For the Euclidean model 
considered in this paper on the other hand,  it is possible to obtain 
mean field exponents by finite size scaling in a finite region of the parameter space.

Similarly,  $\tau$ also does not show finite size scaling. But exactly at $q_c$, it has a 
peak value which scales with $N$ as $\log(N)$. This  agrees with the result of \cite{ben}.
Interestingly, such $\log(N)$ behavior for $\tau$ is not obtained  at $q=q_c$  for any value of $\delta$ for the
Euclidean network. This is  consistent with the earlier results  
  \cite{cohen2001,cohen2003a} that  for small world networks, when percolation is considered, the giant component that emerges does not 
have a small world geometry.  

\section{Discussions of the results}

Before discussing the results, it is useful to recapitulate briefly the 
studies already made on this network as comparison with these 
earlier works is necessary.

\subsection{Earlier studies on this network: Static properties}

It has been mentioned earlier that there are some results available 
for the static properties of the network. 
A number of studies have been made to investigate the behavior of the
network by 
calculating 
numerically the average shortest path as a function of the network size \cite{klein,blumen,psbkc,mouk}.
Two earlier works \cite{psbkc,mouk} in which the greedy algorithm was used to evaluate the shortest paths, showed contradictory results. The study made with much larger   
networks indicated that the network had finite dimensional behavior for 
$1 \leq \delta <  2$ \cite {mouk} while in \cite{psbkc}, 
it was claimed that up to $\delta = 2$, the small world behavior 
exists. In a more recent work \cite{goswami}, all possible shortest paths were
evaluated numerically using a burning algorithm, and it was again found that the network
retains the  small world  property  up to $\delta =2$ while  the clustering coefficient vanishes 
below $\delta =1$. Such a result was also obtained in \cite{biswas}. 
 However,  one can still argue that it is an effect  of finite sizes \cite{mouk}.
 Simple scaling arguments
suggest that there may be   two transitions occurring at $\delta =1.0$
and $\delta = 2.0$ \cite{biswas}. 

\subsection{Critical phenomena studied earlier on this network}

The Ising model has been studied on the Euclidean network and again in two different studies 
contradictory results were obtained. While in \cite{arnab_ps} it was found that the network behaves as a finite dimensional system for $1 < \delta < 2$, the study with the larger network \cite{chang}
indicated a small world behavior up to $\delta = 2.0$. 
Both studies, however, showed that the transition temperature remained $\delta$ independent 
up to $\delta =1$.

SIR is expected to be related to percolation phenomena as already mentioned. Percolation has been studied 
in this type of network recently \cite{havlin}. The detailed study was made in two dimensions. 
In one dimension, results for two values of $\delta = 1.5$ and $1.75$
showed that the exponents are $\delta$ dependent. 
The  values of the threshold infection probability turn out to be larger than our 
results. This is probably because the networks are constructed in 
slightly different ways and also could be because  site percolation was considered  in \cite{havlin}.

\subsection{Present results and comparisons}

In comparison, the results obtained in the present work
do not apparently lead to any definite conclusions for the
behavior of the network for $1 < \delta < 2$, 
while the mean field behavior  obtained for $\delta < 1$ 
is consistent with the earlier results. 
In the mean field region, one has to use the upper critical dimension
in the scaling relation to extract the exponents just like in the case of Ising model \cite{arnab_ps}.

To analyse the region $1 < \delta < 2$ carefully, the results for the 
static exponents $\tilde  \nu$ and $\beta$ may be considered by comparing to the
exponent values of isotropic percolation in different dimensions. 
$\tilde \nu$ remains a constant in this region while $\beta$ shows a change. 
It may be possible that  
 $\tilde \nu $  remains constant  
with  both $\nu $ and $d$   varying.   However, for percolation, we do not have such invariance of 
$\nu d$ for dimensions 1-6 \cite{stauffer}. This suggests that the dimensionality is actually not changing.
 $\beta$ on the other hand does show a deviation from its mean 
field value (equal to 1) for $\delta 
\geq 1.4$.  However, the deviation is not appreciable; the   
extrapolated value of $\beta$ at $\delta = 2.0$ does not  
reach anywhere close to zero as it should. 
Rather, the value of $\beta$ for $\delta$ even very close to 2.0 corresponds to what one would expect for a 
five dimensional lattice which indicates that in all probability  it has a constant value up to $\delta = 2.0$ but shows 
 small changes in  finite systems  due to the   transition occurring at $\delta = 2.0$.

One can also consider the dynamic exponent
 $\mu $ which  shows a stronger dependence on $\delta$ in the region $1 < \delta < 2$ compared to $\beta$.
Here it may be recalled that even in the small world region,
$\mu$ has a finite indicating $\tau$ has a power law behavour 
 while in the finite mean field case, $\tau$ shows a logarithmic behaviour (sec IV).
Hence in contrast to the static exponents, 
$\mu$ does not reflect the behaviour of the network even in the small
world region.   
Thus conclusions about the network's properties cannot be made on the basis of the 
behavior of $\mu$.

We  have to compare  the present results with those of  static percolation on the same network which is available in 
\cite{havlin}. For the results of the two points reported in this paper, the criticality  appears to
change appreciably as $\delta$ varies between 1 and 2 and does not 
show mean field behavior. The disagreement may be because of several factors: 
the construction of the two networks and the average degree is different, secondly, it could be that the correspondence between
percolation and SIR is not entirely true for the Euclidean network in this region. Such lack of correspondence has been noted in some
earlier works \cite{doro_rmp}. Also, there could be such strong finite size 
effects that trying to compare results from two different studies is not 
useful.

\section{Concluding remarks}

From the discussion of the last section, we arrive at the following conclusion:  
SIR on this network shows mean field behavior up to $\delta = 2.0$ for at least the system sizes considered here.
Evidently therefore, the network does not behave as a finite dimensional
lattice for the region $1 < \delta < 2$ as has been claimed in some earlier works but effect of finite sizes are quite strong in the region $\delta 
\geq 1.4$.

Talking about dimensionality, we would like to add a few words here. It is customary to 
use the upper critical dimension $d$ and express $\tilde \nu = \nu d$ to get the mean field 
critical exponents using finite size scaling in these types of systems \cite{doro_rmp}. Hence one uses $d=4$ for the Ising model and $d=6$ for
the percolation case. Surely one cannot assign an unique dimensionality to the network in this 
way at least in the mean field regime. However, assigning a dimension to the network in the mean field region is perhaps meaningless.

 What happens when mean field results are no longer valid?
Theoretically, it is 
possible that one extracts a dimensionality in between 4 and 6 (which is what we are getting in the present work if we consider the value of $\beta$) where percolation
will not show mean field results but Ising model will. This appears far fetched 
and  compels us to believe that
the entire region $0 < \delta < 2$ has mean field behavior in agreement with \cite{chang}. 
%

The critical point shows variation with $\delta$ in the intermediate region as did the transition 
temperature for the Ising model \cite{arnab_ps,chang}. 
Not only that, the time duration also shows appreciable change in the scaling with the 
system size at criticality here. Apparently, the network having appreciable 
clustering leads to these changes which is felt in the intermediate region 
but overall small world property is retained such that mean field behavior prevails. 

We move over to a more profound question, do we reach any conclusion about what
is the actual behavior of the network for $1 < \delta < 2$? 
Still this issue is not resolved. As far as Ising model criticality and
the SIR process is concerned, the latest results suggest it has a small world behavior.
 However, results of static percolation and some geometric properties 
indicate otherwise. The only possible solution  to justify these earlier results  may be that  at finite sizes, the network shows  a 
small world behavior and deviations are only perceptible at very large length scales.

Acknowledgements: The authors thank Arnab Chatterjee for a careful reading of the manuscript. AK acknowledges financial support from UGC sanction no. UGC/534/JRF(RFSMS).

\end{document}